\documentclass[preprint,superscriptaddress,nofootinbib]{revtex4-1}
\pdfoutput=1

\usepackage{hyperref}
\usepackage{bbm}
\usepackage{amsfonts}
\usepackage{mathrsfs}

\usepackage{amsmath,amssymb}

\usepackage{epsfig}
\usepackage{graphicx}               
\usepackage{url}
\usepackage{hyperref}
\usepackage{float}
\usepackage{pstricks}
\usepackage{color}

\newcommand{\nc}{\newcommand}

\nc{\beq}{\begin{equation}}
\nc{\eeq}{\end{equation}}
\nc{\beqa}{\begin{eqnarray}}
\nc{\eeqa}{\end{eqnarray}}
\nc{\bea}{\begin{eqnarray}}
\nc{\eea}{\end{eqnarray}}
\nc{\barray}{\begin{eqnarray}}
\nc{\earray}{\end{eqnarray}}
\nc{\barrayn}{\begin{eqnarray*}}
\nc{\earrayn}{\end{eqnarray*}}
\nc{\ra}{\rightarrow}
\newcommand{\lsim}{\!\mathrel{\hbox{\rlap{\lower.55ex \hbox{$\sim$}} \kern-.34em \raise.4ex \hbox{$<$}}}}
\newcommand{\gsim}{\!\mathrel{\hbox{\rlap{\lower.55ex \hbox{$\sim$}} \kern-.34em \raise.4ex \hbox{$>$}}}}
\nc{\Tr}{{\rm Tr}}
\nc{\slsh}{\slash\hspace*{-0.22cm}}
\def\eg{{\it e.g.}}

\def\be{\begin{equation}}
\def\ee{\end{equation}}
\def\bea{\begin{eqnarray}}
\def\eea{\end{eqnarray}}
\def\bit{\begin{itemize}}
\def\eit{\end{itemize}}

\newcommand{\Eref}[1]{Eq.~(\ref{#1})}
\newcommand{\Fref}[1]{Fig.~\ref{#1}}
\newcommand{\Sref}[1]{Sec.~\ref{#1}}
\newcommand{\vev}[1]{ \left\langle {#1} \right\rangle }
\newcommand{\abs}[1]{ \left| {#1} \right| }

\nc{\infinity}{\infty}
\nc{\mc}{\mathcal}
\nc{\M}{\mathcal{M}}

\def\eg{{\it e.g.}}

\begin{document}

\title{Probable or Improbable Universe?  Correlating Electroweak Vacuum Instability with the Scale of Inflation}

\author{Anson Hook}
\email{hook@ias.edu}
\affiliation{School of Natural Sciences, Institute for Advanced Study\\ Princeton, NJ 08540, USA}
\author{John Kearney}
\email{jkrny@umich.edu}
\affiliation{Michigan Center for Theoretical Physics, University of Michigan, Ann Arbor, MI 48109, USA}
\author{Bibhushan Shakya}
\email{bshakya@umich.edu}
\affiliation{Michigan Center for Theoretical Physics, University of Michigan, Ann Arbor, MI 48109, USA}
\author{Kathryn M. Zurek}
\email{kzurek@umich.edu}
\affiliation{Michigan Center for Theoretical Physics, University of Michigan, Ann Arbor, MI 48109, USA}

\begin{abstract}

Measurements of the Higgs boson and top quark masses indicate that the Standard Model Higgs potential becomes unstable around $\Lambda_I \sim 10^{11}$ GeV.  This instability is cosmologically relevant since quantum fluctuations during inflation can easily destabilize the electroweak vacuum if the Hubble parameter during inflation is larger than $\Lambda_I$ (as preferred by the recent BICEP2 measurement).  We perform a careful study of the evolution of the Higgs field during inflation, obtaining different results from those currently in the literature. We consider both tunneling via a Coleman-de Luccia or Hawking-Moss instanton, valid when the scale of inflation is below the instability scale, as well as a statistical treatment via the Fokker-Planck equation appropriate in the opposite regime. We show that a better understanding of the post-inflation evolution of the unstable AdS vacuum regions is crucial for determining the eventual fate of the universe. If these AdS regions devour all of space, a universe like ours is indeed extremely unlikely without new physics to stabilize the Higgs potential; however, if these regions crunch, our universe survives, but inflation must last a few e-folds longer to compensate for the lost AdS regions. Lastly, we examine the effects of generic Planck-suppressed corrections to the Higgs potential, which can be sufficient to stabilize the electroweak vacuum during inflation. 

\end{abstract}

\preprint{MCTP-14-10}

\maketitle


\section{Introduction}
\label{sec:introduction}

Recent measurements of the Higgs boson and top quark masses, $m_h \approx 125.7 \text{ GeV}$ \cite{Aad:2013wqa,Chatrchyan:2013mxa,ATLAS-CONF-2013-012,CMS-PAS-HIG-13-001} (see \cite{Giardino:2013bma} for combination) and $m_t = 173.34 \text{ GeV}$ \cite{ATLAS:2014wva}, have important implications for the stability of the electroweak vacuum.  For these values, the Standard Model (SM) Higgs potential develops an instability at scales well below the Planck scale (adapted from \cite{Buttazzo:2013uya}),
\begin{equation}
\log_{10} \frac{\Lambda_I}{\mbox{GeV}} = 11.0 + 1.0 \left(\frac{m_H}{\mbox{ GeV}}-125.7\right) - 1.2\left(\frac{m_t}{\mbox{GeV}}-173.34\right)+0.4 \frac{\alpha_3(m_Z)-0.1184}{0.0007},
\label{LambdaI}
\end{equation}
where $\Lambda_I$ is the scale at which the effective Higgs quartic $\lambda_{\rm eff}$ becomes negative.

If the Higgs potential is indeed unstable, the Higgs field $h$ can quantum mechanically tunnel from the electroweak vacuum to the true (unstable) vacuum at large field values.    The lifetime for this tunneling event exceeds the age of the universe, rendering our universe metastable.   As a result, the existence of the additional vacuum at large Higgs vacuum expectation value (vev) does not appear to preclude the existence of our universe.  

This instability becomes cosmologically relevant, however, if the universe underwent a period of inflation.  During inflation, the large inflationary energy density can drive the Higgs out of the electroweak vacuum, since perturbations in the metric induce fluctuations in $h$ of size 
\begin{equation}
\delta h = \frac{H}{2 \pi}.
\label{eq:quantumfluctuation}
\end{equation}
When the scale of these fluctuations, set by the Hubble parameter $H$, is larger than the instability scale of the Higgs potential, the likelihood that $h$ fluctuates to the unstable region of the potential during inflation will be sizable, even if the Higgs field begins inflation in the electroweak vacuum.

The question of the evolution of the Higgs field during inflation has become particularly important in light of the recent BICEP2 \cite{Ade:2014xna} measurement of the tensor-scalar ratio, $r$, which directly probes the scale of inflation,
\begin{equation}
H^2 = \frac{\pi M_P^2 \Delta_R^2  r}{16},
\end{equation}
where $M_P = G^{-1/2} = 1.22 \times 10^{19} \text{ GeV}$ is the Planck mass and $\Delta_R^2 = 2.21 \times 10^{-9}$ is the observed amplitude of the (nearly) Gaussian curvature perturbations as measured by Planck \cite{Ade:2013zuv}.  This translates to a scale of inflation 
\begin{equation}
H \simeq 1.0 \times 10^{14} \mbox{ GeV} \left(\frac{r}{0.16}\right)^{1/2}.
\end{equation}
If the BICEP2 result $r \approx 0.2$ holds, then the Hubble scale is indeed much larger than the Higgs instability scale $\Lambda_I$ over much of the preferred $(m_h, m_t)$ parameter space.  The main analysis and conclusions of this paper, however, are relevant regardless of whether the BICEP2 result holds.

\begin{figure}
\centering
\includegraphics[width=1.0\linewidth]{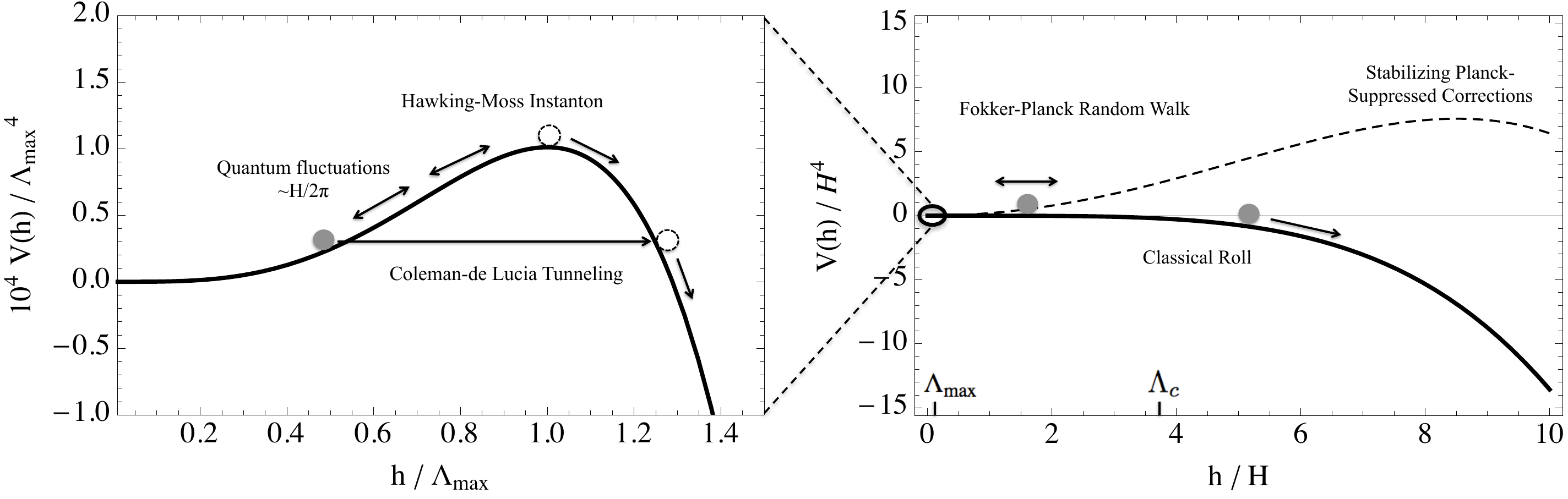} 
\caption{\label{fig:Veff} The Higgs potential illustrated with the regimes of validity for various solutions for the Higgs vacuum evolution during inflation: Coleman-de Luccia (CdL), Hawking-Moss (HM) and Fokker-Planck (FP). {\em Left:} For $H\lsim{\Lambda_{\rm max}}$, the CdL tunneling or single bounce HM instanton yields the transition probability. {\em Right:} For $H\gg{\Lambda_{\rm max}}$, the potential barrier at ${\Lambda_{\rm max}}$ is irrelevant, and a stochastic random walk approach is necessary until classical slow roll takes over at $h = \Lambda_c$ ($H=10\, {\Lambda_{\rm max}}$ has been chosen). The dashed curve in the right-hand panel shows the effect of Planck-suppressed stabilizing terms on the potential (in this case $\Delta V = 0.2 H^2 h^2$), which we study in \Sref{sec:corrections}.  To illustrate the relative scale between the two panels, the dashed lines show the region where the left panel fits into the right panel.}
\end{figure}

To determine the implications of inflation for the stability of the electroweak vacuum, one must follow the evolution of the Higgs through the production of $e^{3 N_e}$ distinct Hubble volumes, where $N_e$ is the number of e-foldings during inflation, and then map this evolution to a probability that a given Hubble volume is in the stable or unstable vacuum at the end of inflation.  As we will show, to correctly follow the probability distribution requires different approaches depending on the relative size of $H$ and the scale $\Lambda_{\rm max}$ where the Higgs potential is maximized (in practice, $\Lambda_{\rm max}$ is not very different from the instability scale $\Lambda_I$ in \Eref{LambdaI}).  The different regimes of the calculations are shown pictorially in \Fref{fig:Veff} and summarized as follows (a more detailed discussion appears in \Sref{sec:higgsevolution}): 
\begin{itemize}
\item When $H \ll \Lambda_{\rm max}$, the potential barrier is classically impenetrable and the vacuum transition proceeds via Coleman-de Luccia (CdL) bubble nucleation \cite{Coleman:1980aw}.
\item As $H$ approaches $\Lambda_{\rm max}$, quantum fluctuations in $h$ of the size \Eref{eq:quantumfluctuation} may drive the Higgs to near the maximum of its potential, at which point the Higgs field will classically roll down the potential.  As a result, the transition from the false vacuum to the true vacuum is dominated by the Hawking-Moss (HM) instanton \cite{Hawking:1981fz}.
\item When the Higgs quantum fluctuations are comparable to or larger than the height of the potential barrier, the fluctuations may drive the Higgs field back and forth over the barrier multiple times during inflation.  The exponential suppression associated with CdL or HM is completely lifted, and the potential barrier at $\Lambda_{\rm max}$ essentially becomes irrelevant.  In this case a statistical treatment via the Fokker-Planck (FP) equation is appropriate. 
\end{itemize}
The Higgs evolution was first studied within the FP approach in Ref.~\cite{Espinosa:2007qp}, while Ref.~\cite{Kobakhidze:2013tn} employed a HM solution.  While we concretely connect our results to these references, we differ in both the understanding of the domain of validity and the implementation of the solutions, as we discuss in more detail in \Sref{sec:higgsevolution}.

Once the probability distribution of the Higgs expectation value has been computed, the next important question is its implication for the evolution of the universe.  The HM or FP probabilities give the distribution of vacua across the $e^{3N_e}$ causally disconnected Hubble patches at the end of inflation.  In the case that $H \lsim \Lambda_{\rm max}$, most of these Hubble patches will be in the safe electroweak vacuum while, when $H \gsim \Lambda_{\rm max}$, most of the Hubble patches are in the unstable vacuum.  The probability that we evolve into a universe that looks like ours depends on the evolution of the unstable vacuum patches once inflation ends.   These regions exhibit a large negative vacuum energy density, so will eventually transition to an anti-de Sitter (AdS) phase and ``crunch.''  However, as they are at a lower energy density than the electroweak vacuum regions, the crunching bubbles of true vacuum can also  ``eat'' the false electroweak vacuum regions. Depending on the relative rates of these processes, unstable regions could either disappear or critically threaten the existence of our universe.  As we shall see, these two outcomes have very different implications in terms of constraints on the scale of inflation $H$.  Ref.~\cite{Espinosa:2007qp} implicitly assumed that the AdS volumes benignly crunch without destroying the stable electroweak vacua, while Ref.~\cite{Kobakhidze:2013tn} did not consider the post-inflationary evolution.  In \Sref{sec:postinflation}, we discuss these scenarios and define different probabilities of the universe surviving depending on the evolution of the AdS vacua. 

In \Sref{sec:corrections}, we discuss corrections to the Higgs potential that can be important during inflation.  We find that Planck-suppressed operators (which one generically expects to be present) can significantly alter the Higgs potential (see the dashed curve in \Fref{fig:Veff}), greatly enhancing electroweak vacuum stability.  Finally, in \Sref{sec:conclusions}, we conclude and identify the outstanding questions for future work.

\section{Higgs Field Evolution During Inflation}
\label{sec:higgsevolution}

\begin{figure}
\centering
\includegraphics[width=3in, height=2.3in]{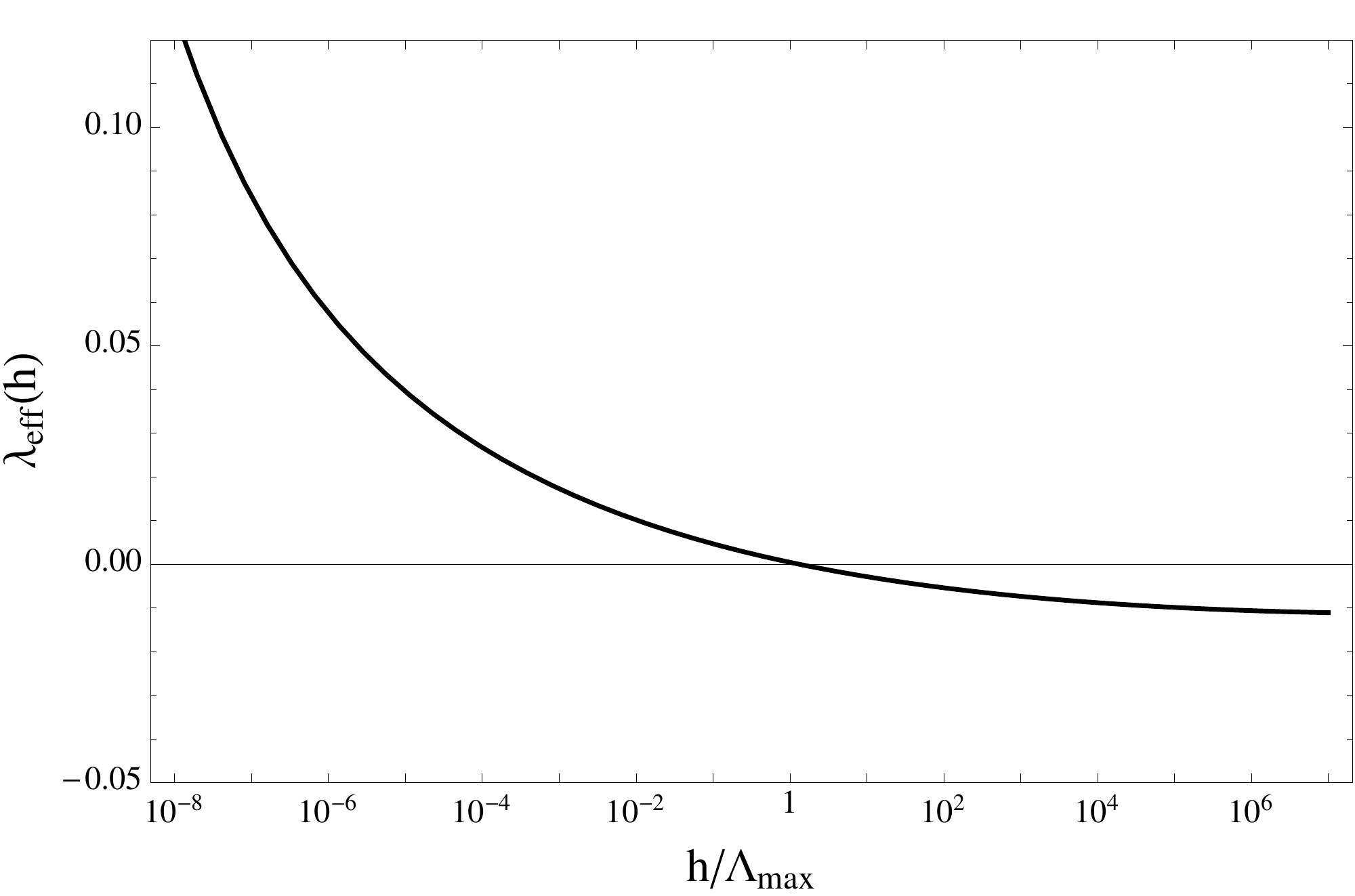} \hspace{0.03\linewidth}
\includegraphics[width=3in, height=2.2in]{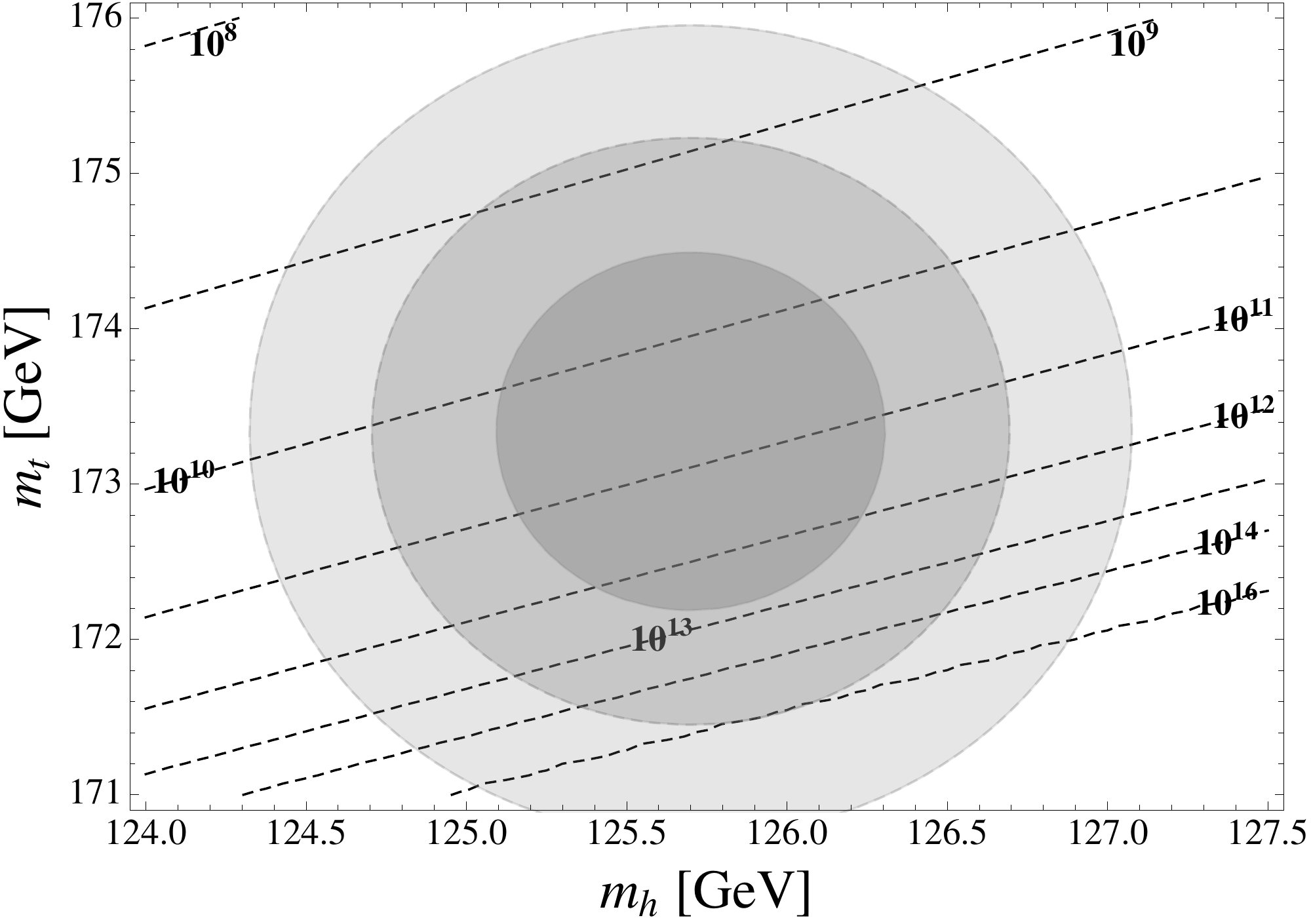} 
\caption{\label{fig:mhmtplot} {\em Left}: $\lambda_{\rm eff}(h)$ within the Standard Model for $m_h = 125.7 \text{ GeV}, m_t = 173.34 \text{ GeV}$. {\em Right}: Contours of $\Lambda_{\rm max}$ (black, dashed) in the $(m_h, m_t)$ plane.  Also shown are ellipses corresponding to the 68.27\%, 95.45\% and 99.73\% confidence level regions for two parameters.  The measured values for the masses are taken to be $m_h = 125.7 \pm 0.4 \text{ GeV}$ and $m_t = 173.34 \pm 0.76 \text{ GeV}$.  For the central values, $\Lambda_{\rm max} = 4.9 \times 10^{10} \text{ GeV}$.}
\end{figure}

In this section, we describe the formalism for studying the evolution of the Higgs field during inflation.  We begin with a single Hubble patch, assuming $\vev{h} = 0$ initially, and follow the Higgs field evolution as this region inflates. Our goal is to calculate the probability that the universe can undergo the necessary amount of inflation without quantum fluctuations knocking the Higgs out of its false vacuum. For large Higgs field values $h \gg v$, where $v = 246$ GeV is the Higgs vev in the electroweak vacuum, we make use of the potential\footnote{Throughout this paper, we employ two-loop renormalization group equations with boundary conditions at $\mu = m_t$ as given in \cite{Buttazzo:2013uya}.  In addition, as in \cite{Buttazzo:2013uya}, we include anomalous dimension and one-loop effective potential contributions to the effective quartic $\lambda_{\text{eff}}(h)$.}
\begin{equation}
V_{\text{eff}}(h) = \frac{\lambda_{\text{eff}}(h)}{4} h^4.
\end{equation}
Since $\lambda_{\text{eff}}$ runs negative at higher scales, the Higgs potential turns over at some scale $\Lambda_{\rm max}$. We show the behavior of $\lambda_{\text{eff}}$ in the left panel of \Fref{fig:mhmtplot}, and $\Lambda_{\rm max}$ in the $(m_h,m_t)$ plane, as well as ellipses corresponding to the 68.27\%, 95.45\% and 99.73\% confidence level regions for the two parameters, in the right panel.
The shape and scale of this potential determine the transition between the three different regimes of CdL, HM, and FP vacuum transitions shown in \Fref{fig:Veff}. Our goal in this section is to explore the Higgs evolution in and elucidate the phenomenological relevance of these regimes.

For simplicity and ease of comparison with earlier studies, we first concentrate on the Higgs potential without any corrections from higher dimension operators; in \Sref{sec:corrections}, we will consider Planck-suppressed corrections to the Higgs potential, which can be significant.  We also assume that $H$ is (to a very good approximation) constant during inflation, in order to study the general phenomenon of electroweak vacuum stability during inflation in a model-independent manner.

\subsection{Various Approaches to Fluctuations Past the Potential Barrier}
\label{subsec:variousapproaches}

Tunneling through a classically impenetrable barrier is calculated using the Coleman-de Luccia (CdL) formalism \cite{Coleman:1980aw}, which gives the nucleation rate of bubbles of true vacuum in a region of false vacuum.  An illuminating interpretation of the CdL transition in de Sitter (dS) space is in a thermal context, as thermally-assisted tunneling \cite{Brown:2007sd}, where the field is thermally excited partially up the barrier and then tunnels through. The thermal contribution is attributed to the existence of an event horizon in dS space, which effectively gives rise to a dS temperature via the Gibbons-Hawking effect \cite{Gibbons:1977mu}
\beq
T_{dS}=\frac{H}{2\pi},
\eeq
where $H$ is the Hubble parameter; correspondingly, the thermal system is defined on a static patch of size $H^{-1}$.

In the semiclassical approximation, the CdL tunneling rate in a unit volume is
\begin{align}
\Gamma & \sim Ae^{-B}, & B & = S_E(\phi) - S_E(\phi_{fv}),
\end{align}
where $S_E(\phi)$ and $S_E(\phi_{fv})$ are the Euclidean actions for the bounce solutions and the false vacuum respectively. The exact calculation of the prefactor $A$ is difficult though its effects are usually subdominant in comparison to the exponential. Estimates of the tunneling probability generally employ the thin wall approximation (see, \emph{e.g.}, \cite{Huang:2008jr}). However, the SM Higgs potential has no true minimum, just a runaway to large field values, such that the thin wall approximation does not apply in this case; a bubble of ``true" vacuum would have the Higgs field changing spatially as well as temporally throughout the bubble, and a solution has to be obtained numerically.\footnote{See \cite{HenryTye:2008xu} for an illuminating discussion of the various cases of CdL tunneling.  For $H \ll \Lambda_{\rm max}$, tunneling through the Higgs barrier corresponds to Case (II) in that paper.}

If the Higgs potential barrier is sufficiently broad, $\abs{V^{\prime \prime}_{\rm eff}(\Lambda_{\rm max})} < 4 H^2$, the CdL instanton does not necessarily exist \cite{Batra:2006rz}, and the transition instead is given by the HM instanton \cite{Hawking:1981fz}. In the thermal picture \cite{Brown:2007sd}, this corresponds to the scenario where the broad potential barrier suppresses the tunneling process and the field is instead thermally excited all the way to the top of the barrier, after which it classically rolls down to its true vacuum with unit probability. The study in \cite{HenryTye:2008xu} found that, in the case of HM tunneling, the inside of the bubble does not reach the true vacuum while the outside of the bubble is cut off by the finite dS horizon before the false vacuum is reached. As such, this transition should be interpreted as an entire Hubble-sized region tunneling to the top of the barrier \cite{Batra:2006rz}. It should be kept in mind that it is only a Hubble patch, and not the entire universe (as had been originally assumed), that makes this transition. The HM transition probability of a Hubble patch from the false vacuum to the top of the barrier during one Hubble time is given by $p \sim e^{-B_{\rm HM}}$ where \cite{Hawking:1981fz, Kobakhidze:2013tn}
\beq
\label{eq:HMform}
B_{\rm HM} = \frac{8\pi^2}{3}\frac{\Delta V}{H^4},
\eeq
with $\Delta V = V_{\rm eff}(\Lambda_{\rm max})$, the height of the Higgs potential barrier relative to the false (electroweak) vacuum. Note that for $H^4 \gg \Delta V$, the exponential suppression factor essentially becomes unity and the prefactor (as well as subleading corrections) becomes important \cite{HenryTye:2008xu}. 

In the $\abs{V^{\prime \prime}_{\rm eff}(\Lambda_{\rm max})} < 4 H^2$ regime, where the transitions are dominated by thermal fluctuations, a stochastic approach to field evolution \cite{Linde:1991sk, starobinsky:stochastic} is more relevant. This involves replacing the quantum fluctuations with a random noise term and studying the ensuing Brownian motion of the field, described by the Fokker-Planck (FP) equation. The FP approach becomes necessary because it captures dynamics that the HM instanton cannot.  Recall that the HM instanton is computed by obtaining the bounce for evolving from the bottom of potential to the top --- it computes a single transition across the barrier. 
When $H \gg \Lambda_{\rm max}$, by contrast, multiple transitions across the barrier are possible.
Moreover, the dynamics of the regions where the field value is larger than $\Lambda_{\rm max}$ are important in determining the final distribution of the Higgs field values at the end of inflation; this information is contained in the FP equation but not in HM. The stochastic approach is therefore necessary in this regime.
However, for $H \lsim \Lambda_{\rm max}$, a single transition across the barrier dominates, and one can show that in this limit the FP equation gives rise to the HM transition probability as expected \cite{Espinosa:2007qp, Starobinsky:1994bd}.

To summarize, the various regimes are: 
\begin{itemize}
\item CdL bubble nucleation is the dominant contribution when $$H^2 \lsim V_{\rm eff}^{\prime \prime}(\Lambda_{\rm max})\sim \lambda_{\rm eff}(\Lambda_{\rm max}) \Lambda_{\rm max}^2.$$
\item A single HM instanton is the dominant contribution when $$V_{\rm eff}^{\prime \prime}(\Lambda_{\rm max}) \lsim H^2 \lsim \left(V_{\rm eff}(\Lambda_{\rm max})\right)^{1/2} \sim \left(\lambda_{\rm eff}(\Lambda_{\rm max})\right)^{1/2} \Lambda_{\rm max}^2.$$
\item FP statistical treatment is needed when the potential barrier becomes small in comparison to the quantum fluctuations, {\em i.e.} for $H \gsim \left(V_{\rm eff}(\Lambda_{\rm max})\right)^{1/4}$.
\end{itemize}

The recent LHC and BICEP2 data suggest that $\abs{V^{\prime \prime}_{\rm eff}(\Lambda_{\rm max})} < 4 H^2$ and likely $H \gsim \left(V_{\rm eff}(\Lambda_{\rm max})\right)^{1/4}$, such that the HM and stochastic approaches are most relevant to Higgs evolution during inflation.  However, even without input from BICEP2, this regime is of far greater interest than the CdL regime.  For $H$ sufficiently small that CdL tunneling dominates, the transition probability is sufficiently suppressed that the likelihood of our universe existing is exponentially close to unity regardless of the evolution of the unstable vacuum patches --- the fluctuations are simply too weak to knock the Higgs out of the electroweak vacuum.  For this reason, we will use the HM solution and the stochastic approach to study the evolution of the Higgs field.  We find that $\lambda_{\rm eff}(\Lambda_{\rm max}) \sim 10^{-4}$, such that the FP regime corresponds to $H/\Lambda_{\rm max} \gsim 0.1$.
In the next subsection, we describe the implementation of the stochastic approach using the Fokker-Planck equation.

\subsection{The Fokker-Planck Equation}
\label{subsec:FPeq}

The probability $P=P(h,t)$ to find the Higgs field at value $h$ at time $t$ satisfies the Fokker-Planck equation \cite{Linde:1991sk, starobinsky:stochastic} 
\beq
\label{Eq:FP}
\frac{\partial P}{\partial t} = \frac{\partial}{\partial h}\left[\frac{V'(h)}{3H} P + \frac{H^3}{8 \pi^2}\frac{\partial P}{\partial h} \right].
\eeq
The first moment of the Higgs field in a time $\tau$ is determined by the equations of motion assuming ``slow roll'' evolution of the Higgs field, $\frac{\vev{\Delta h}}{\tau} = - \frac{V^\prime}{3 H^2}$; this approximation is valid as long as $h \lsim H \sqrt{3/\lambda_{\rm eff}(h)}$. The second moment is dominated by the random fluctuations of the Higgs field, which are driven by the inflationary energy density.  Specifically, in a time $\tau \sim H^{-1}$, the field fluctuates an amount $h \sim \frac{H}{2 \pi}$. We can view this as a random walk with time intervals $\tau \sim H^{-1}$ and step size $\frac{H}{2 \pi}$, such that $\frac{\vev{\abs{\Delta h}^2}}{\tau} = \frac{H^3}{4 \pi^2}$.\footnote{Strictly, this is only true for constant $H$; for time-varying $H(t)$, scalar field variances receive additional corrections.  However, provided the rate of change of the Hubble parameter is small, $\dot{H} \ll H^2$, these additional contributions are negligible such that the Fokker-Planck equation remains applicable in the quasi-de Sitter background (see, \emph{e.g.}, \cite{Starobinsky:1994bd}). Here, we assume $\dot{H} \ll H^2$ as for, \eg, slow-roll models.  We have analyzed the effect of a changing Hubble parameter in several specific models of inflation, and find that the results are numerically equivalent to changing the constant $H$ by a factor less than 2, which will not significantly affect our conclusions.} Since the FP equation has no spatial derivative, the approach is equivalent to studying the evolution of the Higgs field locally at a ``point," where the point is a region of spatial dimension $H^{-1}$ \cite{starobinsky:stochastic}; this is consistent with the observation made earlier that the HM transition should be interpreted as a coherent transition of an entire Hubble-sized region of space. 

The precise solution to the FP equation depends on the initial conditions. We assume that the Higgs field is initially localized at $h=0$, $P(h,0)=\delta(h)$.  Since the typical size of the fluctuation in one Hubble time is $\sim \frac{H}{2\pi}$, the final survival probability after several e-folds of inflation is not very sensitive to this initial choice. For our numerical studies, we employ a sharply-peaked Gaussian distribution for $P(h,0)$ --- our results are insensitive to the exact shape of the Gaussian provided the standard deviation $\sigma \ll \Lambda_{\rm max}$.

In principle, the FP equation is valid for all values of $(h,t)$ provided the slow-roll approximation holds for the Higgs field.  As such, one could simply evolve the initial distribution and use the resulting $P(h,t)$ to determine the likelihood of $h$ taking a particular value in any Hubble patch. However, for practical computational ease and to avoid regimes in which Higgs slow-roll, and hence \Eref{Eq:FP}, does not hold, it is useful to impose boundary conditions.  If the boundary is set at a sufficiently large field value, the solution will converge and be independent of the exact location of the boundary.

In order to determine suitable boundary conditions, we note that in a Hubble time $\Delta t = H^{-1}$ the classical change in the Higgs field due to the potential is (again, assuming slow-roll)
\beq
\label{eq:classical}
\Delta h_{\rm classical} = \dot{h} \Delta t = - \frac{V^\prime_{\text{eff}}(h)}{3 H^2}.
\eeq
Meanwhile, the typical quantum fluctuation is of size
\beq
\label{eq:quantum}
\delta h_{\rm quantum} = \frac{H}{2\pi}.
\eeq
As such, the Higgs field will begin to roll irreversibly down the potential once $\Delta h_{\rm classical} > \delta h_{\rm quantum}$.  We define $\Lambda_c$ to be the point at which classical motion starts to dominate,
\beq
\label{eq:lambdaceqn}
- V^\prime_{\text{eff}}(\Lambda_c) = \frac{3 H^3}{2 \pi}.
\eeq
Consequently, $P(h \gsim \Lambda_c,t)$ rapidly flattens out as the Higgs field rolls away.  Thus, a suitable approximation to $P(h,t)$ can be achieved, particularly in the regime of interest $\abs{h} \leq \Lambda_{\rm max}$, by employing the boundary condition $P(h = \Lambda_c,t) = 0$.\footnote{While we use this boundary condition for our analytic analysis below, we have verified numerically that the probability distribution in the region of interest does not change as the cutoff is increased.}

In \Fref{fig:mhmtplot} (left panel), we show $\lambda_{\rm eff}(h)$ for $m_h = 125.7 \text{ GeV}, m_t = 173.34 \text{ GeV}$; the effective quartic becomes approximately constant for $h > \Lambda_{\rm max}$, such that we can approximate $V^\prime_{\text{eff}}(h > \Lambda_{\rm max}) \approx \lambda_{\rm eff}(h) h^3$.  Then
\beq
\label{eq:cutoff}
\Lambda_c \approx \left(\frac{3}{2 \pi \lambda_{\rm eff}(\Lambda_c)}\right)^{\frac{1}{3}} H \approx 3.6 \left(\frac{-0.01}{\lambda_{\rm eff}(\Lambda_c)}\right)^{\frac{1}{3}} H.
\eeq
Therefore, quantum fluctuations remain larger than the classical effect until $h \sim \mathcal{O}({\rm few}) H$.  As a consistency check, we note that within this approximation the condition for slow roll $\abs{V^{\prime \prime}_{\rm eff}(h)} \ll 9 H^2$ requires $h \lsim 17.3 \sqrt{-0.01/\lambda_{\rm eff}(h)} H$, such that the aforementioned boundary does indeed avoid the region where slow-roll breaks down.\footnote{Although we show approximate values here, in our numerical studies we determine $\Lambda_c$ by solving \Eref{eq:lambdaceqn}.}  Moreover, since the total energy density is dominated by the inflaton energy density until $V_{\rm eff}(h) \sim -V(\phi)$ (which requires the Higgs roll off to $\abs{h} \sim \sqrt{H M_P}$), inflation proceeds unabated for $\abs{h} < \Lambda_c$.  As such the FP equation, which models dS-to-dS transitions, remains valid in this regime.

An alternative boundary condition previously considered in the literature \cite{Espinosa:2007qp} is $P(\abs{h} = \Lambda_{\rm max},t) = 0$.  This is an appropriate boundary condition if the field rolls down to the true minimum with unit probability once it fluctuates to the top of the barrier.  However, this is not the case for large quantum fluctuations $H \gsim \left(V_{\rm eff}(\Lambda_{\rm max})\right)^{1/4}$ -- such a boundary condition artificially forces $P(h,t)$ to vanish at a particular point $h = \Lambda_{\rm max}$ determined by the potential in spite of the fact that the classical motion due to the potential is negligible in this region.  Consequently, the resulting solution underestimates the probability distribution in the regime $h \leq \Lambda_{\rm max}$.  Analogously, in this regime, it is insufficient to consider the transition probability due to a single HM instanton.  For $H \gsim \left(V_{\rm eff}(\Lambda_{\rm max})\right)^{1/4}$, the HM instanton has an order one probability, hence multiple HM-like transitions back and forth over the barrier occur over the course of inflation.  Setting $P(\abs{h} = \Lambda_{\rm max},t) = 0$ neglects the large probability of fluctuating back over the barrier.

As an aside, we comment that this also has interesting implications for the tuning of the initial configuration of the Higgs field at the onset of inflation. If the Higgs field can take on any value below $M_P$ at the start of inflation and the Higgs must start within its shallow potential well if the electroweak vacuum is to be realized in some regions of space, this would require a fine-tuning of the initial conditions at the level of $\Lambda_{\rm max}/M_P$. However, in light of the above picture, it appears possible to start with the Higgs field value on the order of the Hubble scale and still realize the electroweak vacuum due to quantum fluctuations \textit{into} the false vacuum, relaxing the amount of  tuning required to $\sim H/M_P$. Since the combination of LHC and BICEP2 data suggests $H \gg \Lambda_{\rm max}$, this is a significant improvement in tuning by several orders of magnitude.

At the end of inflation, Hubble patches where the field has fluctuated to regions beyond the top of the barrier will roll off to the AdS vacuum (we assume that reheating does not modify the Higgs potential sufficiently to push these field configurations back to the electroweak vacuum). In the remaining Hubble patches, the Higgs field will roll back to the electroweak or Minkowski vacuum. The probability of landing in the electroweak vacuum at the end of inflation is therefore given by 
\beq
\label{eq:plambdadef}
P_{\Lambda}\equiv\int_{-\Lambda_{\rm max}}^{\Lambda_{\rm max}}dh\, P(h,t_e)
\eeq
where $t_e$ denotes the end of inflation. $P_{\Lambda}$ should be interpreted as the probability of a region that is of Hubble size towards the end of inflation to land in the electroweak vacuum. Note that we have not included the effects of thermal fluctuations in the Higgs during the reheating phase at the end of inflation. If the reheat temperature is high enough, these effects can further destabilize the electroweak vacuum; this issue is well addressed in \cite{Espinosa:2007qp}.

\subsection{Approximate Solution to the Fokker-Planck Equation}
\label{subsec:approxFP}

For $H^3 \gg V_{\rm eff}^\prime(\Lambda_{\rm max})$, the FP equation admits an approximate analytic solution. In this limit, the effect of the Higgs potential barrier as well as the Higgs contribution to the Hubble parameter are negligible, and the first term on the right-hand side of \Eref{Eq:FP} can be dropped provided $\Delta h_{\text{classical}} < \delta h_{\text{quantum}}$.  In the opposite regime $\Delta h_{\text{classical}} > \delta h_{\text{quantum}}$, the Higgs potential eventually causes the field to rapidly roll off to large vevs. This is (to a good approximation) taken into account by the boundary condition $P(\Lambda_c,t)=0$.

With this boundary condition and the initial condition $P(h,0)=\delta(h)$, one obtains the following approximate solution\footnote{The derivation is the same as that in \cite{Espinosa:2007qp}, except for the modified boundary condition.}
\begin{align}
\label{survival2}
P_\Lambda & = \frac{2}{\pi} \sum_{n = 0}^{\infty} \frac{1}{n + \frac{1}{2}} \exp\left\{-\left(n + \frac{1}{2}\right)^2 \int\frac{H^3}{8 \Lambda_c^2}dt\right\} \sin\left(\left(n + \frac{1}{2}\right) \frac{\pi \Lambda_{\rm max}}{\Lambda_c}\right) \notag \\
& \simeq \frac{2 \Lambda_{\rm max}}{\Lambda_c} \exp\left\{-\frac{N_e}{32} \left(\frac{H}{\Lambda_c}\right)^2\right\} \left(1+\exp\left\{-\frac{N_e}{4} \left(\frac{H}{\Lambda_c}\right)^2\right\} + \exp\left\{-\frac{3N_e}{4} \left(\frac{H}{\Lambda_c}\right)^2\right\} \right) ,
\end{align}
where in the second line we have retained the leading terms in the sum and taken $\Lambda_c \gg \Lambda_{\rm max}$.  We also take $H$ to be approximately constant during inflation such that $\int H dt=N_e$, where $N_e$ is the number of e-folds for which inflation occurs.

The leading exponential reflects the fact that the survival probability falls exponentially with the duration of inflation, as one might expect given that a longer period of inflation allows more time for the probability distribution to spread out due to the quantum fluctuations.\footnote{Using \Eref{eq:cutoff}, we note that the leading exponent is $\sim N_e/300$.  As such, the transition rate out of the false vacuum is too slow to terminate inflation globally.}  Meanwhile, as the potential is negligible for $h \lsim \Lambda_c$, $P(h,t)$ rapidly becomes broad relative to $\Lambda_{\rm max}$.  The result is that $P(\abs{h} < \Lambda_{\rm max},t)$ is essentially flat; this is reflected by the $\Lambda_{\rm max}/\Lambda_c$ factor.

As mentioned in the preceding subsection, boundary conditions are employed for simplicity only and our numerical results for $P_\Lambda$ are independent of the exact location of the boundary.  This may seem inconsistent with \Eref{survival2}, which depends on $\Lambda_c$.  However, assuming $\Lambda_c \gsim H \gsim \Lambda_{\rm max}$, this series converges to very similar values even for different choices of $\Lambda_c$, and achieving good agreement with the numerical results simply requires sufficient terms are kept in the sum.  Truncation at three terms is suitable for $\Lambda_c$ as defined by \Eref{eq:lambdaceqn}, $N_e \approx 50-60$ and the SM Higgs potential -- for a higher cutoff values $\Lambda_c^\prime > \Lambda_c$, more terms would have to be kept to achieve the same level of agreement with the numerical results as the exponential terms in parentheses are less suppressed.

\begin{figure}
\centering
\includegraphics[width=0.8\linewidth]{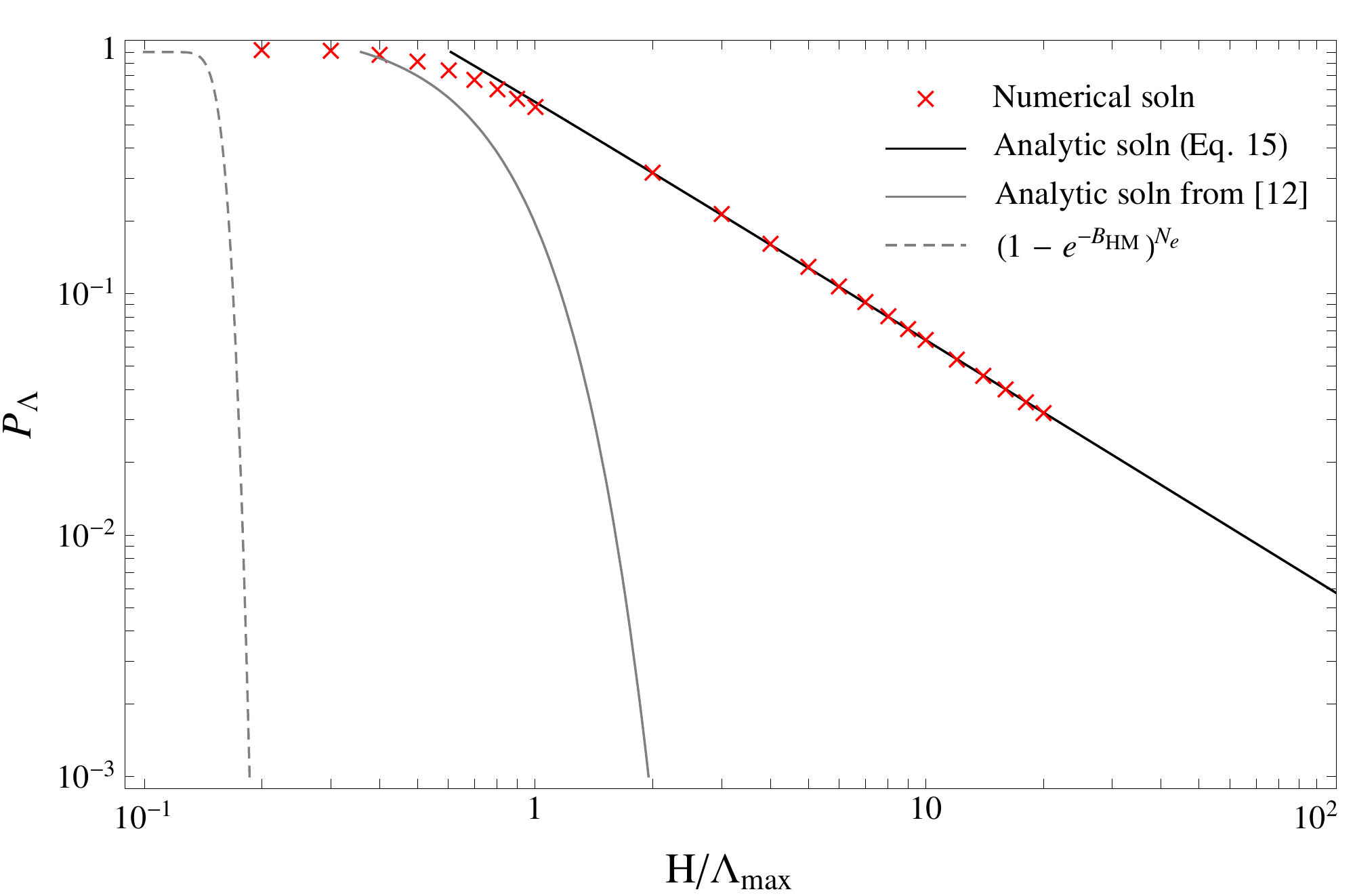} 
\caption{\label{fig:survcomparison} Comparison of survival probability $P_\Lambda$ (\Eref{eq:plambdadef}) with $N_e = 60$ as given by numerically solving the FP equation (red crosses), the approximate analytic solution in \Eref{survival2} (solid black curve), and the solution with $\Lambda_c=\Lambda_{\rm max}$ from \cite{Espinosa:2007qp} (solid gray curve).  For comparison, we also show the HM probability as explored in \cite{Kobakhidze:2013tn} (with unit prefactor, see \Eref{eq:HMform}).  $V_{\rm eff}(h)$ in our numerical solution is computed using the central values $m_h = 125.7 \text{ GeV}$, $m_t = 173.34 \text{ GeV}$.}
\end{figure}

In \Fref{fig:survcomparison}, we show our numerically evaluated FP solutions (red crosses) and the analytic form of \Eref{survival2} (solid black curve).  The numerical results were obtained using the boundary condition $P(h = \Lambda_c,t) = 0$, but we have verified that $P_\Lambda$ is unchanged when a higher cutoff is employed.  Given the excellent agreement between the analytic and numerically evaluated solutions, we will henceforth use \Eref{survival2} for our calculations in the $H/\Lambda_{\rm max} \geq 1$ regime. This result should be compared with that of Ref.~\cite{Espinosa:2007qp}, in which the boundary condition was set at $\abs{h} = \Lambda_{\rm max}$ instead of $\Lambda_c$, resulting in a significantly smaller survival probability $P_\Lambda \sim \exp\left\{-\frac{N_e}{32} \left(\frac{H}{\Lambda_{\rm max}}\right)^2\right\}$. The difference between these results is clearly visible in \Fref{fig:survcomparison}, where this survival probability (the solid gray curve) dips sharply for $H/\Lambda_{\rm max} \gsim 1$, whereas the probability from our FP calculation dies off more gradually.

\section{Post-Inflation Evolution and the Fate of the universe}
\label{sec:postinflation}

The previous section provided the necessary tools to track the evolution of the Higgs field during inflation, as a single Hubble-sized region inflates to $e^{3 N_e}$ Hubble regions that are causally disconnected from each other.  In the regions that exit inflation with $\abs{h} < \Lambda_{\rm max}$, $h$ rolls down to the electroweak vacuum, forming regions of Minkowski space.  In the regions with $\abs{h} \geq \Lambda_{\rm max}$, the Higgs rolls off to arbitrarily large field values and arbitrarily negative potential energy, causing these regions to become AdS. As inflation ends and reheating begins, the horizon starts to expand again.  Previously causally disconnected regions of space, some of which are in the acceptable electroweak vacuum and some of which are in the crunching AdS vacuum, come back into causal contact.  What happens as this merging process occurs --- whether the asymptotically AdS vacuum regions crunch or expand to eat the good electroweak vacua --- determines the eventual fate of the universe.

Analyzing the evolution of these patches during this epoch is not a trivial task.  
Ref.~\cite{Freivogel:2007fx} demonstrated that, when AdS and Minkowski bubbles collide in a dS background, whether or not the AdS bubble is expelled depends on the tension in the domain wall separating the bubbles relative to the energy density in the AdS region.
While qualitative aspects of this picture apply to our study, Ref.~\cite{Freivogel:2007fx} only considered empty bubbles and studied their evolution in the thin wall approximation.  In our case, the thin wall approximation fails because the true minimum of the Higgs potential, if it exists, lies far below the false minimum.
Moreover, neither the AdS nor the Minkowski bubbles are empty; the negative energy of the unstable Higgs vacuum can come to dominate the energy density only once the Higgs has rolled out to sufficiently large vevs or once inflation has ended.  In either case, different phases of matter from both the inflaton (during reheating) and Higgs (due to the combined effect of its potential plus kinetic energy) are present and affect the bubble evolution. Even during inflation, the Higgs field and the inflaton evolve differently in distinct Hubble patches (a smaller Hubble parameter due to a significantly negative Higgs potential energy would cause the inflaton to roll down its potential faster), implying that different patches will exit inflation at different times, further complicating the picture. Finally, knowledge of the true vacuum of the Higgs potential, which remains unknown, is necessary to determine the behavior of the AdS regions.

It is impossible to determine the exact fate of the universe without detailed understanding of these aspects --- in light of the complications, we postpone this study for future work \cite{workinprogress}.  However, we can determine the consequences of a given scenario for the existence of our universe.  In particular, the two extreme possibilities are:
\begin{enumerate}
\item All AdS regions crunch rapidly before the domain walls can expand out and take over the Minkowski vacua.  The Minkowski regions survive while the AdS regions vanish. 
\item Any AdS domain wall moves out and takes over all of Minkowski space.  The AdS regions dominate the universe, ultimately resulting in a big crunch.  In this case, a single AdS volume in the past light cone of the observable universe is disastrous. 
\end{enumerate}
We now study the implications for the likelihood of the existence of our universe in each of these two extreme cases.

\subsection{AdS Regions Crunch}
\label{subsec:AdScrunch}

We consider first the optimistic case that AdS regions simply crunch without destroying neighboring Minkowski regions. 
The approximate analytic form in \Eref{survival2} gives the probability for a Hubble-sized region at the end of $N_e$ e-folds of inflation to survive quantum fluctuations and remain within the shallow Higgs potential barrier for $H \gsim \Lambda_{\rm max}$. Meanwhile, the physical volume of space grows as $e^{3N_e}$. Therefore the volume of space (measured via the number of e-folds $N_o$ that survive) after the AdS regions crunch is given by
\beq
e^{3 N_o} = e^{3N_e}P_\Lambda=\frac{2 \Lambda_{\rm max}}{\Lambda_c} \exp\left\{3N_e-\frac{N_e}{32} \left(\frac{H}{\Lambda_c}\right)^2\right\} \left(1 + \ldots \right).
\label{eq:survivingvol}
\eeq
This quantity grows with $N_e$ (for $\Lambda_c = 3.6 H$, the term in the exponent is $3N_e - 0.0024 N_e$). Therefore, although the survival probability of a single Hubble region decreases as inflation is prolonged, the physical volume of surviving space grows; fluctuations into AdS regions effectively slow down the rate of inflation, but cannot completely counter the exponential growth.  

\begin{figure}
\centering
\includegraphics[width=3in, height=2.2in]{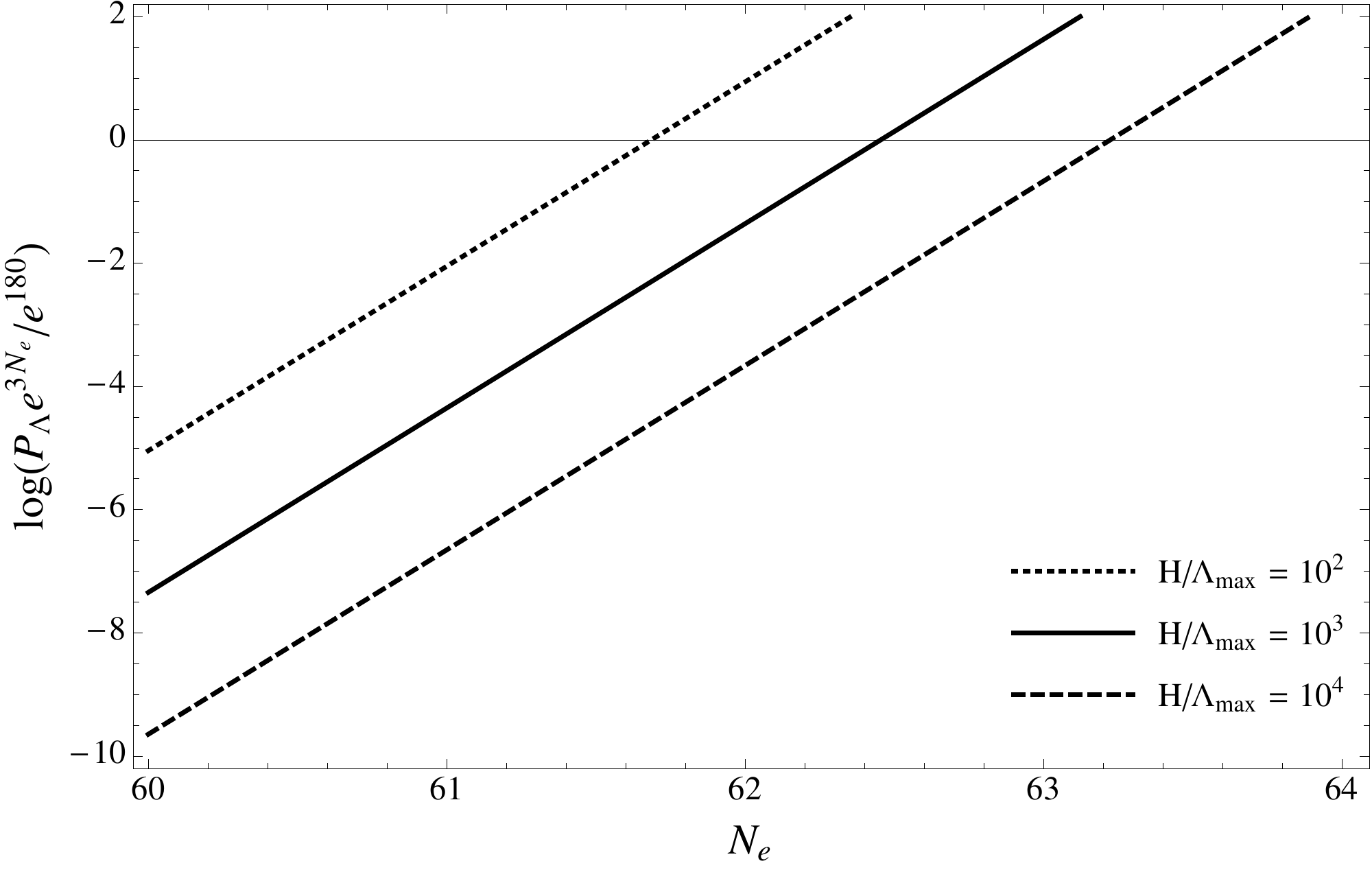} \hspace{0.03\linewidth}
\includegraphics[width=3in, height=2.3in]{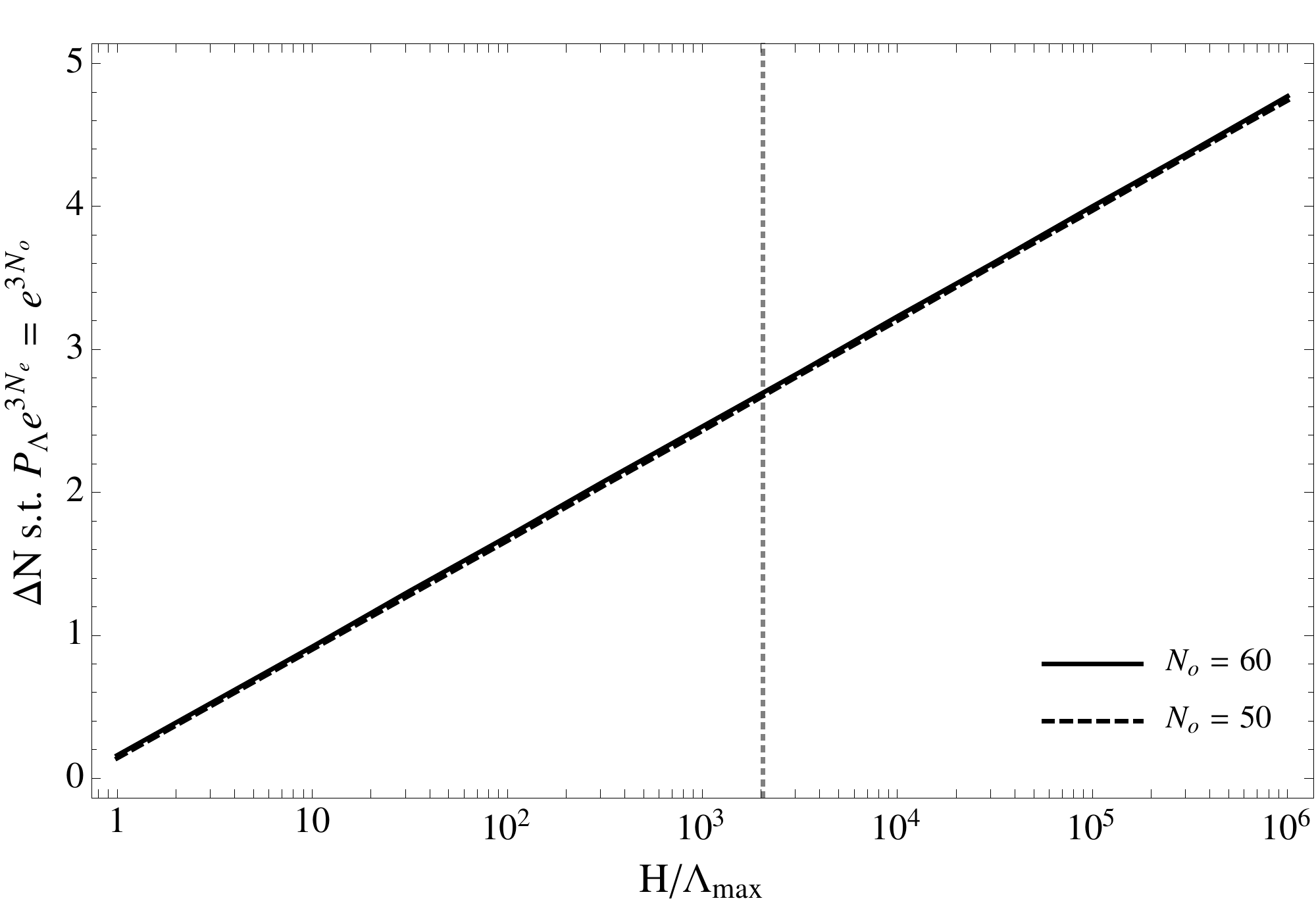} 
\caption{{\em Left:} Total volume of the surviving region (normalized to $e^{180}$ Hubble volumes) after $N_e$ e-folds of inflation for different values of $H/\Lambda_{\rm max}$. {\em Right:} Additional number of e-folds of inflation needed in order to obtain $e^{N_o}$ Hubble volumes of space left over after the AdS regions crunch.  $\Lambda_c$ is determined by solving \Eref{eq:lambdaceqn} with the Higgs potential for $m_h = 125.7 \text{ GeV}$, $m_t = 173.34 \text{ GeV}$.  The gray, dotted line corresponds to $H_{\rm BICEP2} = 10^{14}$ GeV.}
\label{fig:survol3}
\end{figure}

Since $N_o \gsim 50-60$ to solve the homogeneity and flatness problems, inflation must occur for a longer period $N_e > N_o$ to compensate for volumes that crunch. The left panel of \Fref{fig:survol3} shows the total volume of the surviving region (normalized to $e^{3 N_o} = e^{180}$ Hubble volumes) after $N_e$ e-folds of inflation for various ratios $H/\Lambda_{\rm max}$ with $m_h$ and $m_t$ fixed to their central values.
One can also compute the additional number of e-folds, $\Delta N = N_e - N_o$, required to produce the desired $e^{3 N_o}$ surviving Hubble volumes for a particular value of $H/\Lambda_{\rm max}$ --- the result is shown in the right panel of \Fref{fig:survol3}.
Requiring $e^{3 N_o} = e^{3N_e}P_\Lambda$ implies
\beq
\Delta N = - \frac{1}{3} \log\left(\frac{2 \Lambda_{\rm max}}{\Lambda_c}\right) + \frac{N_e}{96} \left(\frac{H}{\Lambda_c}\right)^2 - \frac{1}{3} \log(1+\ldots).
\eeq
The survival probability, and hence $\Delta N$, is determined predominantly by the ratio $\Lambda_{\rm max}/\Lambda_c$ --- corrections due to different $N_o = N_e - \Delta N$ are subdominant for $N_o \sim 50-60$ (as is clear from the coincidence of the two lines in \Fref{fig:survol3}).  Furthermore, the relatively mild dependence on the hierarchy between $H$ and $\Lambda_{\rm max}$ can be traced to linear dependence of $P_\Lambda$ on $\Lambda_{\rm max}/\Lambda_c \sim \Lambda_{\rm max}/H$ (see \Eref{survival2}).  This is related to the observation of \Sref{sec:higgsevolution} that $\Lambda_{\rm max}$ is only relevant at the end of inflation, since for $H \gsim \Lambda_{\rm max}$ the potential does not significantly influence the evolution of the Higgs field until $h \sim \Lambda_c$.
As a result, even when the Hubble parameter is several orders of magnitude larger than $\Lambda_{\rm max}$, the number of Hubble regions necessary to give the observable universe can be obtained with only a few more e-folds of inflation, $\Delta N \lsim 5$.  It is also clear that, for $H \lsim \Lambda_{\rm max}$, the survival probability is sufficiently close to unity that $\Delta N \approx 0$.

Our analysis and results paint a qualitatively different picture to that in \cite{Espinosa:2007qp, Kobakhidze:2013tn}. In these studies, the rate of formation of AdS regions was always found to be significantly greater than the rate of inflation for $H \gg \Lambda_{\rm max}$ (see, \emph{e.g.}, the gray curves in \Fref{fig:survcomparison}).  Consequently, the universe is unlikely to inflate sufficiently --- the probability of achieving $e^{3N_o}$ surviving Hubble volumes from a single Hubble patch is always vanishingly small and a universe like ours is extremely disfavored. In contrast, we find that the increase in surviving Hubble patches with each e-fold of inflation is sufficient to overcome the decrease due to patches transitioning to the AdS regime.

One can ask whether the crunching AdS regions leave behind any trace of their existence. Ref.~\cite{Espinosa:2007qp} asserted that the AdS regions simply crunch and vanish without trace, prompting a study of the size of curvature perturbations in an anthropic window to obtain a meaningful measure of the probability of landing in a universe like ours.\footnote{A similar study can also be performed within our framework; this, however, is inflation-model-dependent and not directly relevant to the current paper, so we postpone it for future work.}  However, the eventual fate of crunching AdS regions is not clear. Ref.~\cite{Freivogel:2007fx}, for instance, suggests that a crunching AdS bubble surrounded by Minkowski space probably forms an ordinary spherical black hole. If this is true, the collapsing AdS regions should litter our universe with primordial black holes.  While such black holes are likely light enough to have evaporated away fairly rapidly, this is nonetheless an interesting line of inquiry that merits further study. 

\subsection{AdS Regions Dominate}

We next consider the opposite limit, a pessimistic scenario in which a single AdS patch in our past light cone is sufficient to destroy the asymptotically Minkowski regions and hence our universe.   
Since our past light cone must contain at least $e^{3N_o}$ Hubble patches, for our universe to have a non-negligible survival probability the transition probability for a single Hubble patch must be exponentially close to unity, requiring $H \ll \Lambda_{\rm max}$.  
In addition, as $\lambda_{\text{eff}}(\Lambda_{\rm max})$ is small, the potential barrier tends to be broad and sufficiently high in comparison to $H$ that we can compute the transition probability using a HM calculation.  We find $\lambda_{\text{eff}}(\Lambda_{\rm max}) \sim 10^{-4}$ and $\abs{V^{\prime \prime}(\Lambda_{\rm max})} \sim 10^{-3} \Lambda_{\rm max}^2$, such that $\abs{V^{\prime \prime}(\Lambda_{\rm max})} < 4 H^2$ for
\begin{equation}
\frac{H}{\Lambda_{\rm max}} \gsim 10^{-2}.
\end{equation}

As mentioned previously in \Eref{eq:HMform}, the HM probability for a single Hubble patch to fluctuate out of the false vacuum during one Hubble time is $p$, where $p \sim e^{-B_{\text{HM}}}$ and 
\begin{equation}
\label{eq:BHM}
B_{\text{HM}} = \frac{8 \pi^2 \Delta V}{3 H^4} = \frac{2 \pi^2 \lambda_{\text{eff}}(\Lambda_{\rm max})}{3} \left(\frac{\Lambda_{\rm max}}{H}\right)^4
\end{equation}
For $p\ll 1$, one can approximate the survival probability during a single Hubble time as $1-p\approx e^{-p}$. Thus, the probability of no Hubble patches transitioning during inflation to the destructive AdS regime in our past light cone is
\begin{equation}
\label{eq:pnoads}
P_{\rm no AdS} \sim \prod_{N_e = 1}^{N_o} (e^{-p})^{e^{3 N_e}} = \prod_{N_e = 1}^{N_o} e^{-e^{3N_i - B_{\text{HM}}}}
\end{equation}
and
\begin{equation}
-\log P_{\rm no AdS} \sim \sum_{N_e = 1}^{N_o} e^{3N_e - B_{\text{HM}}} = e^{-B_{\text{HM}}} \frac{e^3 (e^{3N_o} - 1)}{e^3 - 1} \simeq e^{3 N_o - B_{\text{HM}}},
\end{equation}
where the last approximate equality indicates that the probability is dominated by the final e-fold of inflation, as one would expect from exponential growth.

\begin{figure}
\centering
\includegraphics[width=0.45\linewidth]{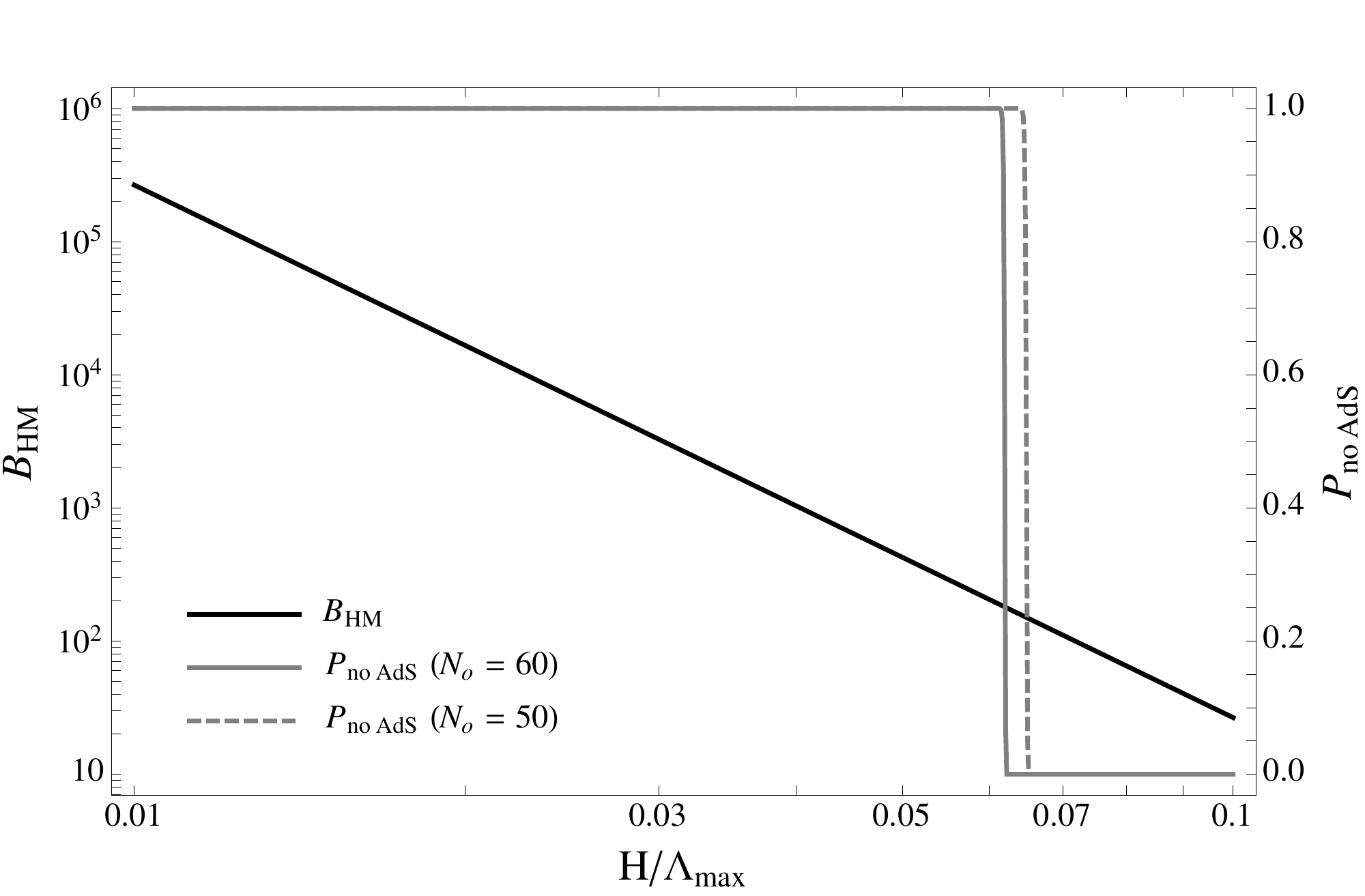} \hspace{0.03\linewidth}
\includegraphics[width=0.45\linewidth]{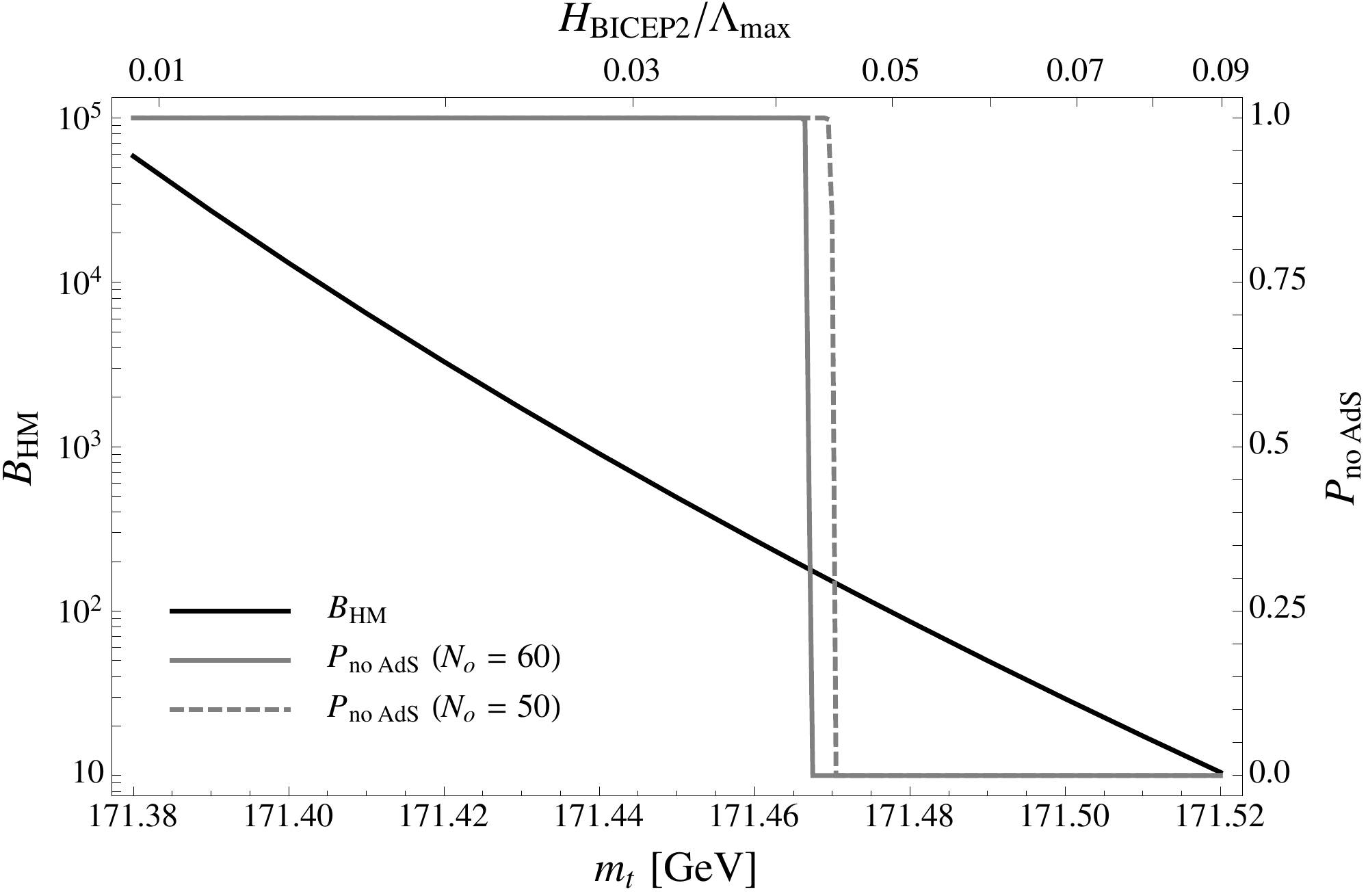} 
\caption{\label{fig:BHMandPnoAdS} $B_{\rm HM}$ and $P_{\rm no AdS}$ as a function of $H/\Lambda_{\rm max}$ for the central values $m_h = 125.7 \text{ GeV}$, $m_t = 173.34 \text{ GeV}$ (left) and as a function of $m_t$ for $m_h = 125.7 \text{ GeV}$ and $H_{\rm BICEP2} \approx 10^{14} \text{ GeV}$ (right).  The survival of our universe either requires $H/\Lambda_{\rm max} \lsim 0.065$ or, if the BICEP2 result holds, $m_t \lsim 171.47 \text{ GeV}$, $\sim 2.5 \sigma$ below the central value.}
\end{figure}

\begin{figure}
\centering
\includegraphics[width=0.6\linewidth]{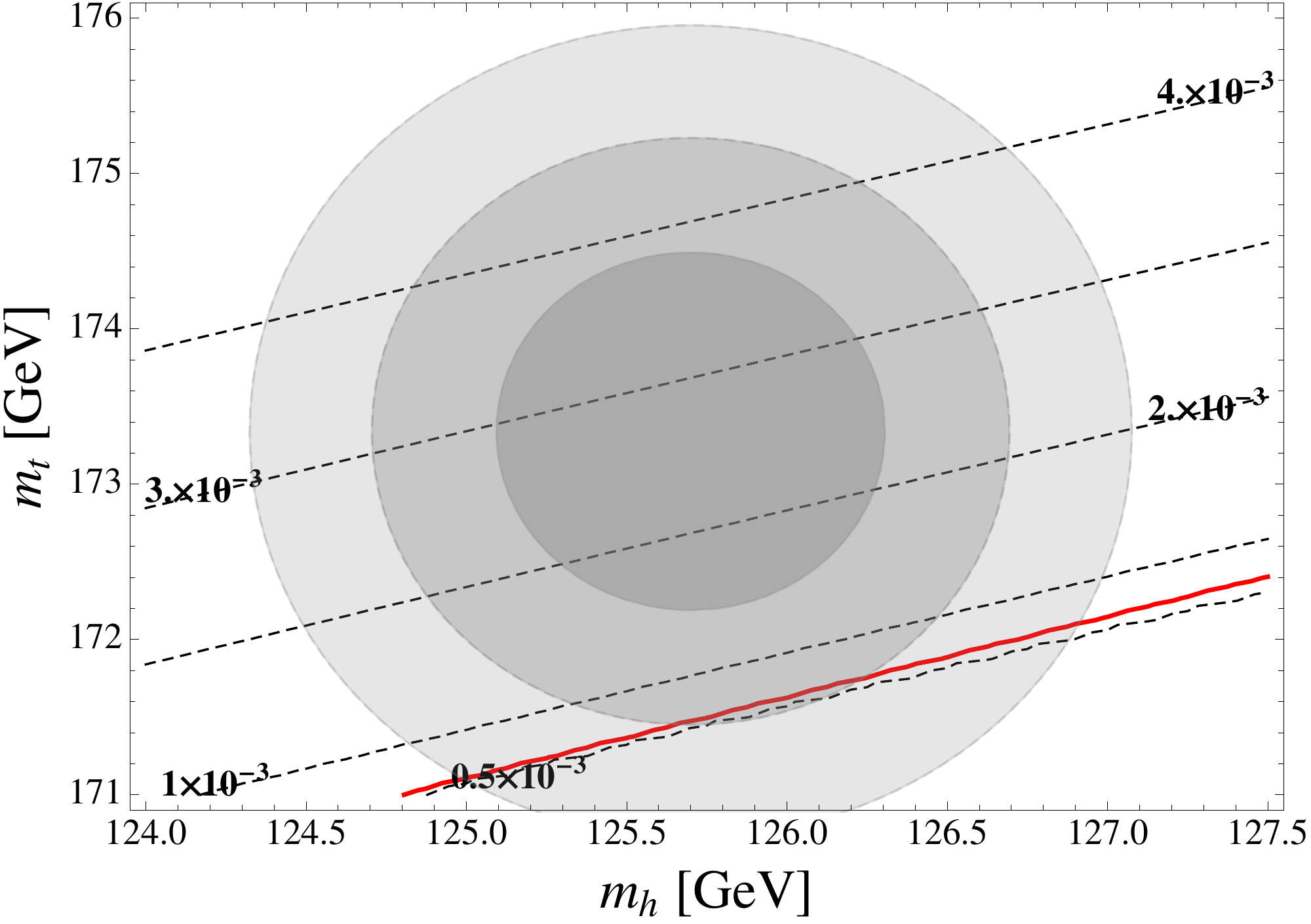} 
\caption{\label{fig:mhmtplotBHM} Contours of $\frac{2 \pi^2 \lambda_{\rm eff}(\Lambda_{\rm max})}{3}$ (black, dashed) in the $(m_h, m_t)$ plane, which can be used to determine the approximate limit on $H/\Lambda_{\rm max}$ from \Eref{eq:HLambdamaxapproxlimit}.  The red line corresponds to $B_{\rm HM} = 3 N_o = 180$ for $H_{\rm BICEP2} \approx 10^{14} \text{ GeV}$ ({\em i.e.}, the region exhibiting a significant survival probability lies below this line).}
\end{figure}

Achieving a non-negligible survival probability $P_{\rm no AdS} \gsim e^{-1}$ requires $B_{\text{HM}} \gsim 3 N_o$ or, from \Eref{eq:BHM},
\begin{equation}
\label{eq:HLambdamaxapproxlimit}
\frac{H}{\Lambda_{\rm max}} \lsim \left(\frac{2 \pi^2 \lambda_{\text{eff}}(\Lambda_{\rm max})}{9 N_o}\right)^{1/4} \sim \mathcal{O}(0.1).
\end{equation}
The exact limit on $H/\Lambda_{\rm max}$ depends on $\lambda_{\text{eff}}(\Lambda_{\rm max})$, which is determined by the boundary values for the couplings at $\mu = m_t$ (and hence by, \emph{e.g.}, $m_h, m_t$).  However, as the limit goes as the fourth root of $\lambda_{\text{eff}}(\Lambda_{\rm max})$, it does not vary significantly throughout the preferred parameter space.  In \Fref{fig:BHMandPnoAdS}, we show $B_{\text{HM}}$ and $P_{\rm no AdS}$ as a function of $H/\Lambda_{\rm max}$ for the central values $m_h = 125.7 \text{ GeV}$, $m_t = 173.34 \text{ GeV}$, and as a function of $m_t$ with $H$ fixed to the approximate BICEP2 value $H_{\rm BICEP2} \approx 10^{14} \text{ GeV}$.  As expected, the survival probability becomes negligible almost instantaneously once $B_{\text{HM}} \lsim 200$, corresponding to $H/\Lambda_{\rm max} \gsim 0.065$ (for $m_t = 173.34 \text{ GeV}$) or $m_t \gsim 171.47 \text{ GeV}$ (for $H = 10^{14} \text{ GeV}$).  In \Fref{fig:mhmtplotBHM}, we show contours of $\frac{2 \pi^2 \lambda_{\text{eff}}(\Lambda_{\rm max})}{3}$, which can be used to determine the limit on $H/\Lambda_{\rm max}$ for a particular choice of $(m_h, m_t)$, and delineate the region of parameter space exhibiting a significant survival probability for $H = H_{\rm BICEP2}$ and $N_e = 60$.  

Given the approximate form of \Eref{eq:HMform} and the large powers involved in \Eref{eq:pnoads}, we would like to stress that the results here should be taken as estimates of where the survival of the universe switches from being improbable to likely.

\section{Corrections to the Higgs Potential During Inflation}
\label{sec:corrections}

To this point, we have considered a purely SM Higgs potential, valid up to the Planck scale.  However new physics, such as new states associated with the solution to the hierarchy problem (\emph{e.g.} supersymmetry), can modify the Higgs potential and may even render the electroweak vacuum stable.  Even in the absence of new states, we expect the Higgs potential to be modified by finite temperature and gravitational corrections.  Thermal corrections to the Higgs potential are negligible during inflation but gravitational corrections, as we will show, are crucially important.  In this section, we discuss possible gravitational corrections and the implications for the stability of the electroweak vacuum.

Quantum gravity is expected to generate higher dimension operators in the Higgs potential, suppressed by powers of $M_P$.  Despite this suppression, these operators may be relevant for electroweak vacuum stability for two reasons.  First, large $h$ values may counter some of the $M_P$ suppression, leading to non-negligible contributions to the potential as $h$ rolls away.  Consequently, these corrections may stabilize the runaway direction.  Second, the large inflationary vacuum energy density due to the inflaton, $V_I(\phi) \sim M_P^2 H^2$, can make higher dimension operators important during inflation.  Specifically, as gravity couples to energy, the Higgs-graviton couplings may generate a sizable effective mass for the Higgs via a ``gravitational Higgs mechanism,'' which significantly alters the shape of the Higgs potential.

Consider the following dimension six Planck-suppressed corrections to the Higgs potential,
\begin{equation}
\Delta V = \frac{\alpha}{M_P^2} V_I(\phi) h^2 + \frac{\alpha_2}{M_P^2} h^6.
\end{equation}
As $V_I(\phi) \sim H^2 M_P^2$, the first term generates a mass for the Higgs field set by $H$ that, if $\alpha > 0$, serves to stabilize the Higgs potential to scales $h \sim H$.\footnote{For quadratic inflation, this is analogous to the stabilizing term considered in, \emph{e.g.}, \cite{Lebedev:2012sy}.}  Since the transition from a highly probable to an improbable universe (with the requirement that there be no unstable Hubble patches in the past light cone) happens when $H \sim \Lambda_{\rm max}$, this term is crucially important for the evolution of the Higgs during inflation.  The second term stabilizes the runaway direction of the Higgs potential for $\alpha_2 > 0$, leading to a true minimum at $h \lsim M_P$.  This affects the energy density in the true vacuum patches --- $h$ settles down to the minimum of the potential, as opposed to continuously rolling to larger field values --- and so could ultimately influence the evolution of these patches, determining which of the scenarios discussed in \Sref{sec:postinflation} (true vacuum patches crunch or dominate) is realized.\footnote{One might have thought that the dimension eight operator $\frac{\alpha_3}{M_P^4} V_I(\phi) h^4$ could potentially counter the negative Higgs quartic coupling and prevent the Higgs potential from turning over at all for $H \lsim M_P$.  However, even for the large $H$ favored by BICEP2, $V_I(\phi)/M_P^4\sim H^2/M_P^2\sim10^{-10}$.  The coefficient of the quartic term in the Higgs potential is $\abs{\lambda_{\rm eff}/4} \sim 10^{-3}$ at large energies, so this correction is irrelevant for reasonable values of $\alpha_3$.}

To examine the effects of the first operator, we parameterize the Higgs potential as
\begin{equation}
\label{eq:VPlanckSlop}
V(h) = \frac{c}{2} H^2 h^2 + \frac{\lambda_{\text{eff}}(h)}{4} h^4
\end{equation}
and investigate the implications for electroweak vacuum stability, focusing on the stabilizing choice $c > 0$.  Neglecting the running of $\lambda_{\text{eff}}(h)$ (which is small for large $h$, see \Fref{fig:mhmtplot}), the potential is maximized for
\begin{equation}
\frac{\Lambda_{\rm max}}{H} = \sqrt{\frac{c}{\abs{\lambda_{\text{eff}}(\Lambda_{\rm max})}}}
\end{equation}
such that, for $\lambda_{\rm eff} \approx -0.01$ and $c \sim \mathcal{O}(1)$ (a reasonable estimate from an effective field theory perspective), $\Lambda_{\rm max} \sim \mathcal{O}(10) H$.  In this case, the probability for a single Hubble patch to remain in the electroweak vacuum during inflation becomes significant (as in \Fref{fig:survcomparison} with $\Lambda_{\rm max} \gsim H$).  A HM calculation gives the transition probability provided $0.1 \lsim c \lsim 2$ --- for smaller $c$, $H \sim \Lambda_{\rm max}$ such that you enter the Fokker-Planck regime, whereas for larger $c$ the potential barrier is no longer sufficiently broad that the HM instanton dominates.  Again neglecting the running of $\lambda_{\text{eff}}(h)$,
\begin{equation}
B_{\text{HM}} \approx \frac{2 \pi^2 c^2}{3 \abs{\lambda_{\text{eff}}(\Lambda_{\rm max})}} \approx 660 c^2,
\end{equation}
where the final approximate equality uses $\abs{\lambda_{\text{eff}}(\Lambda_{\rm max})} \approx 0.01$ (as in \Fref{fig:mhmtplot} for large $h$).  Thus, $B_{\rm HM} \gsim 3 N_o = 180$ requires $c \gsim 0.5$.

\begin{figure}
\centering
\includegraphics[width=0.45\linewidth]{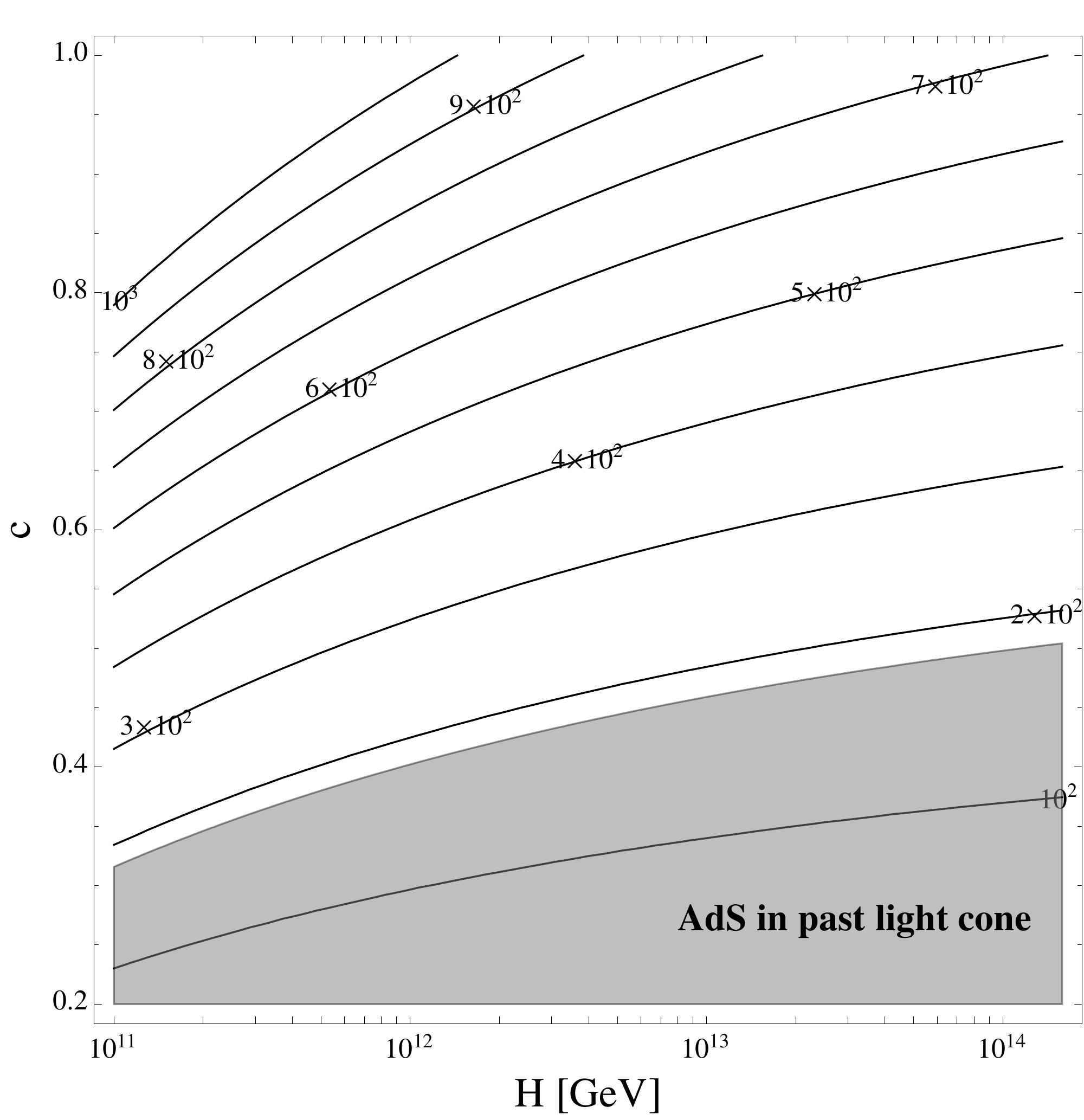} \hspace{0.03\linewidth}
\includegraphics[width=0.45\linewidth]{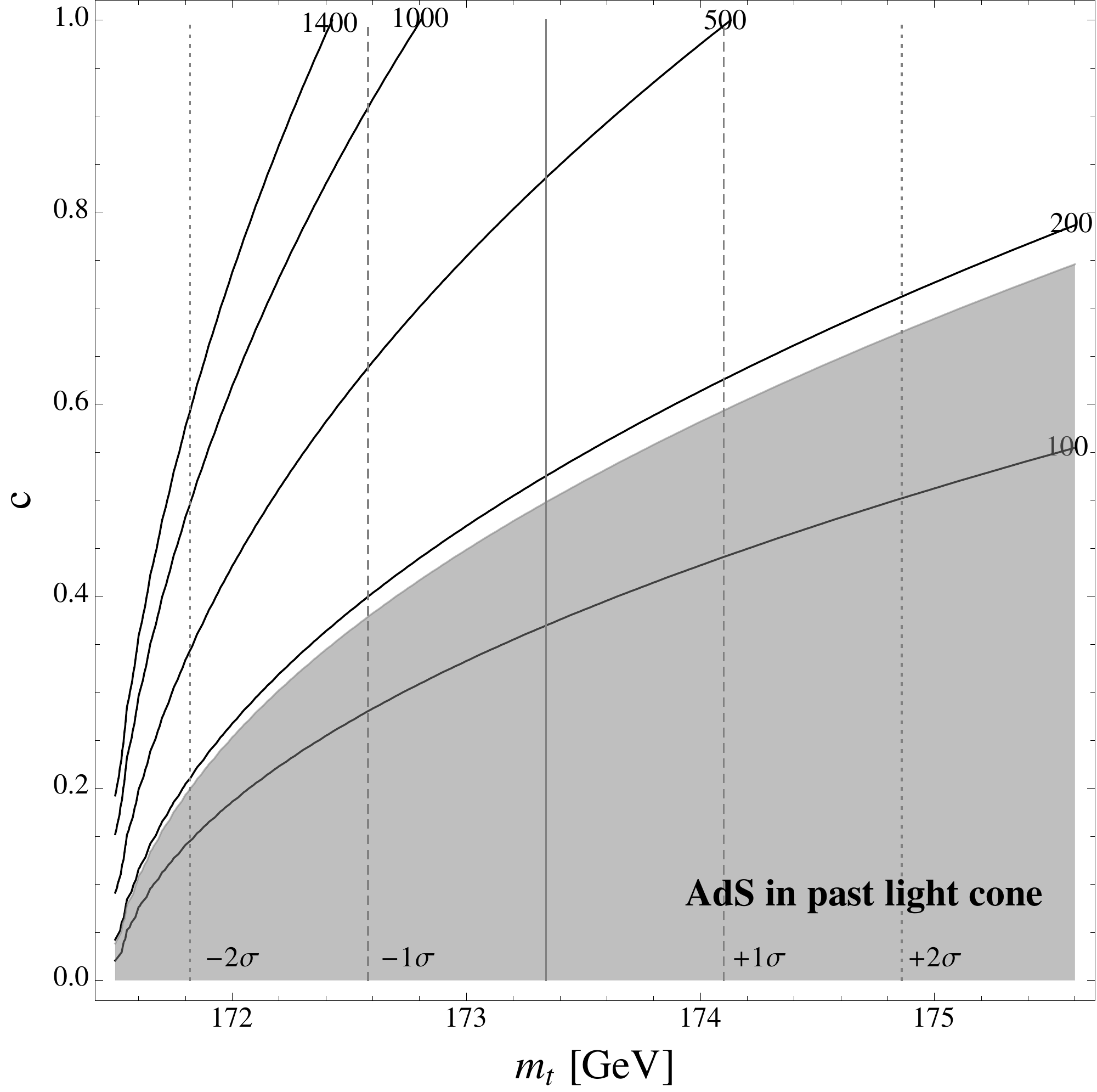} 
\caption{\label{fig:BHMHc} Contours of $B_{\rm HM}$ as a function of $c$ and $H$ for the central values $m_h = 125.7 \text{ GeV}$, $m_t = 173.34 \text{ GeV}$ (left) or $m_t$ for $m_h = 125.7 \text{ GeV}$ and $H_{\rm BICEP2} \approx 10^{14} \text{ GeV}$ (right).  The gray shaded region denotes $B_{\rm HM} < 180$, in which case the survival probability of our universe becomes negligible when a single AdS volume in our past light cone would destroy our universe.  In the right plot, we also show the $2\sigma$ regions for $m_t$.  Over much of the parameter space, a modest $c \lsim \mathcal{O}(1)$ is sufficient to stabilize the potential and thus increase the likelihood of our universe surviving inflation.}
\end{figure}

The actual values of $\Lambda_{\rm max}$ and $B_{\text{HM}}$, and hence of $c$ necessary to sufficiently stabilize the potential, depend on the running and precise value of $\lambda_{\text{eff}}$.  This is taken into account in our numerical studies, but the above are reasonable estimates in the regime of interest.  In \Fref{fig:BHMHc}, we show contours of $B_{\text{HM}}$ as a function of $H$ and $c$ for the central values $m_h = 125.7 \text{ GeV}$ and $m_t = 173.34 \text{ GeV}$, and as a function of $m_t$ and $c$ for $m_h = 125.7 \text{ GeV}$ and $H_{\rm BICEP2} = 10^{14} \text{ GeV}$.  The gray, shaded region in \Fref{fig:BHMHc} delineates $B_{\text{HM}} < 180$ (such that $P_{\rm no AdS} < e^{-1}$).  For reasonable values of $c$, $B_{\text{HM}}$ is sizable and so the probability for a Hubble patch to remain in the electroweak vacuum is significant.  If true vacuum patches crunch, this does not significantly alter the conclusions of \Sref{subsec:AdScrunch} beyond that the number of e-folds required to yield $e^{3N_o}$ surviving Hubble patches is very close to $N_o$, $\Delta N \approx 0$.  If unstable patches dominate, a moderate value for $c$ can ensure that the probability of a Hubble patch transitioning out of the electroweak vacuum within our past light cone is sufficiently small that our universe is not disfavored even if $H \gg \Lambda_{\rm max}$.

\section{Conclusions and Open Questions}
\label{sec:conclusions}

We have studied the evolution of the Higgs field during inflation in the presence of a potentially catastrophic true minimum.  Our goal was to understand how improbable our universe is given the conditions for inflation that likely existed in the early universe.  This has become especially relevant in light of the BICEP2 results, which favor a scale of inflation $H \simeq 10^{14} \mbox{ GeV}$, likely several orders of magnitude larger than the scale at which the Higgs potential is maximized, $\Lambda_{\rm max}$ (see \Fref{fig:mhmtplot}).  

We focused on elucidating and delineating the appropriate calculation in three different regimes of validity: where Coleman-de Luccia (CdL), Hawking-Moss (HM) or Fokker-Planck (FP) evolution should be applied.  In particular, we presented numerical and analytical solutions to the FP equation that are valid when $H/\Lambda_{\rm max} \gsim 0.1$, as preferred by the combination of LHC measurements and the BICEP2 results.  In this regime, we find that our calculation of the survival probability differs substantially from earlier results in the literature.

We then considered the implications of these probabilities for the fate of our universe.  For the LHC and BICEP2 preferred values for $H$ and $\Lambda_{\rm max}$ we find that, provided the unstable regions of space rapidly crunch, the universe survives (even though the unstable vacuum is exponentially more likely to be populated than the electroweak vacuum) but must have undergone a few more e-folds of inflation to compensate for the crunched regions. If this is indeed the scenario realized in nature, the additional period of inflation must be taken into account along with horizon and flatness considerations in calculating the minimum number of e-folds of inflation required to produce the observed universe. On the other hand, if the unstable vacuum regions come to dominate and destroy the (relatively rare) electroweak vacuum regions, then our universe is extremely improbable (as an exponential of an exponential). Even in the context of a multiverse, such low probabilities call for the existence of new physics to stabilize the electroweak vacuum up to scales that are on the order of the inflationary scale.  However, the new physics does not need to involve, \emph{e.g.}, new particles with couplings to the Higgs field --- as we discussed in \Sref{sec:corrections}, Planck-suppressed operators with sufficiently large ($\mathcal{O}(1)$) coefficients can provide the necessary stabilization.

An important open question is which of the scenarios for the post-inflationary vacuum evolution --- AdS crunch or domination --- is realized.  This question is complicated due to the interplay of the coupled inflaton and Higgs energy densities in the asymptotically Minkowski and AdS regions of space.  We leave this question for future work.

\vspace{1cm}

{\em Note added:} While this work was in preparation, Refs.~\cite{Fairbairn:2014zia, Enqvist:2014bua, Kobakhidze:2014xda} appeared, which have some qualitative overlap with our work in the limit $H \gg \Lambda_{\text max}$ relevant for the BICEP2 results.     

\newpage

{\em Acknowledgments:} We would like to thank Michele Papucci and Aaron Pierce for valuable discussions. J.K. and B.S. are supported by the DoE under contract de-sc0007859.  A.H. is supported by the DoE under contract de-sc0009988.  K.Z. is supported by NSF CAREER award PHY 1049896.

\bibliography{refs}

\end{document}